# Lessons from a Large-Scale Assessment: Results from Conceptual Inventories

Beth Thacker, Hani Dulli, Dave Pattillo, and Keith West
*Physics Department, Texas Tech University, Lubbock TX 79409*

## Abstract

We report the conceptual inventory results of a large-scale assessment project at a large university.  We studied an attempt at introducing materials and instructional methods informed by physics education research (PER-informed materials) into a department where most instruction has been traditional and a significant number of faculty are hesitant, ambivalent or even resistant about the introduction of such reforms. The changes were made in the laboratories and recitation sections of the introductory classes, both calculus-based and algebra-based, introducing PER-informed materials and training the teaching assistants in student-centered instructional methods. In addition to the results found in the large lecture classes, we present the results of a small PER-informed, inquiry-based, laboratory-based class that has been taught as a special section of the algebra-based course for about 10 years. The assessment reported in this paper was done using available PER-developed assessment instruments. The results of other assessment instruments used in the project, such as free-response pre- and post-tests, are reported in subsequent papers. The results in this paper inform researchers in PER of the use of PER-informed materials and instructional methods in a department not unified in the introduction and implementation of these materials and the results of the implementation as assessed by PER-based conceptual inventories. We found that our conceptual inventory scores were lower than many results reported elsewhere in the literature. However, we did see a statistically significant increase in the conceptual inventory scores with the implementation of PER-informed laboratories and the use of student-centered pedagogy in the labs and recitations. The increase was much greater, if the lecture instructor also used PER-informed materials. The results will also be useful to faculty at other similar institutions.

## I. INTRODUCTION

While the introduction and adoption of Physics Education Research-informed (PER-informed) materials[1] and teaching techniques, both by departments and individual faculty, are becoming increasingly more common, there are still barriers to reform[2-5] and changes can be met with significant faculty resistance.[6] There are still institutions where traditional lecture instruction is the norm and any implementation of research-based materials is done by instructors on an individual basis, without departmental concurrence and often without departmental support. Texas Tech University (TTU) is one such university. It is a large research university where the instruction in physics is primarily traditional and there has not been a unified approach to the teaching of the introductory physics courses. There is, however, a small subset of instructors with an interest in reform. In 2007, an undergraduate committee was formed that decided on the introduction of PER-informed materials into the laboratories and the implementation of recitation sections that would include the use of PER-informed materials. With the support of the Department Chair at the time, even though not all faculty were unified or in agreement, PER-informed materials were introduced into the laboratories and the newly formed recitation sections. We had a situation, then, where changes were implemented in the laboratories and recitation sections, but the lecture instruction remained unchanged.

The changes began in Spring 2008, with that semester as a transitional semester, with some PER-informed materials used in the laboratories. In Fall 2008, we began the implementation of PER-informed laboratories in the algebra-based courses and also introduced recitation sections in those courses. In Spring 2009, the implementation of PER-informed laboratories in the calculus-based course was begun and recitation sections were introduced into those courses also. In 2009, we applied for and were awarded (in Fall 2009) a National Institutes of Health (NIH) Challenge grant[7] to support a large-scale assessment of the introductory courses and the changes being made in the laboratories and recitation sections.

We set out to assess all of the introductory physics courses, both calculus-based and algebra-based, using existing assessment instruments. The main research-based assessment instruments in common use are multiple-choice conceptual inventories. There are very few, if any[8], valid and reliable comprehensive assessment instruments, research-based or not, in general use designed explicitly for the university level introductory physics courses. We chose to use four different conceptual inventories: the Force Concept Inventory (FCI),[9] the Brief Electricity and Magnetism Assessment (BEMA),[10] the Mechanics Baseline Test (MBT)[11] and the Conceptual Survey of Electricity and Magnetism (CSEM).[12] While these only assess conceptual change in certain content areas, this information combined with the results from other assessment instruments, gave us an indication of the results of changes we had made.

In addition, these assessment instruments are valid and reliable and the results can be compared to those at other universities. The other assessment instruments we used include Scientific Attitude and Scientific Reasoning Inventories[13], locally written free-response pre- and post-tests, and TA Evaluation Inventories[14]. In this paper, we report only on the results of the administration of the conceptual inventories.  The results of the other assessments will be reported in other papers.

In addition to the changes being made in the large lecture classes, we wished to assess a laboratory-based, inquiry-based course[15-18] that was developed with National Science foundation (NSF) funding about 10 years ago and has been taught as a special section of the algebra-based course every semester since then. It was developed explicitly for health science majors, taking their needs, learning styles, backgrounds and motivations into account. It is taught without a text in a Workshop-Physics style[19] environment and is an inquiry-based course in the manner of <u>Physics by Inquiry</u>,[20] developed by the Physics Education Group at the University of Washington, but at the algebra-based level. The materials were developed by modifying and adapting parts of existing materials designed for other populations and integrating them with new units in our own format, creating a course aimed specifically at health science majors.

There are many papers that report a significant rise in normalized gain when conceptual inventories are used to assess the use of PER-informed materials in the PER literature. So many, that it is easy to forget 1) that simply the introduction of these materials, independent of the institutional environment and contextual factors, does not necessarily produce a significant rise in normalized gain,[21-22] and 2) that conceptual inventories assess only one aspect of understanding and are not meant to be and should not be used as the sole indicator of the success (or failure) of particular materials or pedagogy. Conceptual inventories are not comprehensive assessment instruments and should not be used that way. However, they do reflect the degree to which an intervention applied within certain environmental and contextual factors has affected aspects of conceptual understanding. And it is useful to compare normalized gain results to those at other institutions, when interventions have been applied in similar or different environments and contexts.

So, we present our results from the implementation of PER-informed materials in the labs and recitations, with and without PER-informed materials in the lecture in our specific environment, to add to the collection of such results from the assessment of PER-informed implementations as evidenced by conceptual inventories. We also discuss the need for the use of other assessment instruments to develop a broader and deeper understanding of the effect of curricular changes. In our case, we also have results from free response pre- and post-testing and other assessment instruments that assess different aspects of instruction from the same study. The inclusion of those results in a single paper, would make the paper much too long, and we report those results elsewhere.[23]

Based on the present literature and our knowledge of our own environment, we would predict that the PER-informed laboratory and recitation interventions would have some positive effect on the results of conceptual inventories, but that the effect would vary depending on the lecture instruction. We would also predict that the conceptual inventory results would be higher, if the lecture instructor used PER-informed instruction than if they didn't.

The prediction for the inquiry-based course is a little bit different. We have historical data[16] on the FCI from 6-10 years ago, where the FCI scores were usually in what is known as the interactive-engagement region on a "Hake" plot.[24] However, the scores have always been in the lower end of that region. So, we would predict the scores on the FCI to remain about the same, in the interactive-engagement region, as the course has not been significantly altered over the past few years. We were, however, interested in the results on other conceptual inventories and, more importantly, the results on other forms of assessment. We expected the results, for example, on the free-response pre- and post-tests to be different from the other classes. The goals of the course are not just on altering common pre-conceptions, but include teaching skills such as "thinking like a scientist," demonstrating the ability to apply aspects of critical thinking and solving problems. These other goals are possibly not best assessed by conceptual inventories and change in those areas is more likely to show up on other assessments. We were, however, interested in the results on different conceptual inventories than the FCI and in the comparison of the results on conceptual inventories to those from other classes.

The conceptual inventory assessment results are definitely informative and useful to our own institution, but we believe that it will also be informative to those at other institutions in similar situations. There is evidence that large research institutions are less likely to have adopted PER-informed instructional materials and practices.[4] Our results will inform other institutions of the value of the introduction of PER-informed materials into the laboratories and recitations, with and without PER-informed instruction in the lecture.

In this paper, we discuss, in Section II, the student populations, the state of the introductory courses being assessed, the changes being made to the courses and the teaching styles of the instructors; in Section III, the administration of the assessment instruments; in Section IV, the results; in Section V an analysis of the results and in Section VI, we conclude.

## II. THE DEPARTMENT AND STUDENT POPULATIONS

Texas Tech University (TTU), is a large university of about 32,000 students, with 26,000 of them undergraduates. The physics department has 20 tenured/tenure-track faculty and teaches about 2,600 students in the introductory physics courses each year. This includes the calculus-based and algebra-based introductory physics classes. About 1800 of these students are in the calculus-based course and 800 in the algebra-based course. The introductory courses are usually taught by faculty, but may be taught by postdoctoral researchers, visiting faculty, or even graduate students on occasion.

## A. The laboratories and recitation sections

Prior to Spring 2008, the introductory courses consisted of three hours of lecture and two hours of laboratory work each week. The labs were taught by teaching assistants (TAs). They were very traditional, "cookbook," in format and pedagogy and had not undergone significant change in many decades. The students would work through the labs and turn in a formal lab write-up to the TA. There was no recitation.

After a transitional semester in Spring 2008, PER-informed laboratories and a one-hour weekly recitation section were introduced into the algebra-based courses in Fall 2008. At first, the labs and the recitations were held in three-hour blocks, with two hours of lab and one hour of recitation. This was done mostly to help with scheduling problems, as the recitations were added in. In the first course of the algebra-based sequence (ABI), the Module I of the Real Time Physics labs[25] was used exclusively. In the second course in the sequence (ABII), some Real Time Physics labs  (from Modules 3 and 4) were used and some locally written PER-informed labs were used. By Fall 2010, the ABII labs were almost completely locally written PER-informed labs. They did not require a formal lab write-up, but included laboratory homework. There were also bi-weekly quizzes in the recitation sections that included material from lecture, lab and/or recitations.

Beginning in Spring 2009, a one-hour recitation was also implemented in the calculus-based courses, which we will refer to as Calculus-based I and II (CBI and CBII). The CBI labs used some of the Real Time Physics laboratories and some of the traditional laboratories. The CBII course remained traditional labs. The labs in CBI remained partially Real Time Physics and partially traditional and the labs in CBII remained traditional until Fall 2010. Starting in Fall 2010, up to the present, the labs in the second course in each sequence (ABII and CBII) were almost completely locally written PER-informed labs. The labs in the first course in the calculus-based sequence used Real Time Physics labs exclusively in Fall 2010. After Fall 2010 to the present, labs developed at the University of Illinois[26] were used.

With the introduction of PER-informed labs and recitation sections, the TAs were trained in different pedagogies than were used in the traditional labs. Most of the PER-informed labs were designed in a format that required interactive-engagement (IE) during the lab, to help guide the students.  The labs did not require a formal write-up, but included laboratory homework. There were also bi-weekly quizzes in all of the recitation sections that included material from lecture, lab and/or recitations.

### 1. PER-informed labs

The locally written PER-informed labs consisted of five parts: Objectives, Overview, Explorations, Investigations and Summary. The Objectives listed the concepts and skills the students should understand and be able to demonstrate after completing the lab. The Overview was a short summary of the purpose of the lab. The Explorations were qualitative measurements or, sometimes, qualitative problems or thought experiments designed to focus on concepts the students may still have difficulty with, even after instruction. They might, for example, focus on drawing magnetic field lines or observing the direction compasses point near a current-carrying wire with and without current flowing through the wire.  They allowed students to experimentally observe concepts they had studied and repeat skills and review concepts that would be needed in the Investigation part of the lab.  The Investigation part of the lab consisted of quantitative measurements and observations, taking data, graphing, analyzing and interpreting it. Students would, for example, measure the magnitude of the magnetic field at different distances from a current-carrying wire and plot the data. It is the section that is more like a traditional lab.  In the Summary, the students were asked to focus on a particular part of the lab and summarize it. There was a lab homework to be completed and turned in at the next lab, but no formal lab report. A sample lab is included in Appendix I.

## 2. Recitation sections

The recitation sections were about 50 minutes long and were usually group problem solving sessions monitored by the TA. The problems were chosen by the lab coordinator(s) and were often chosen from or modified versions of published PER-informed problems, such as problems from Tasks Inspired by Physics Education Research (TIPER)[27], Ranking Task Exercises in Physics[28], books by Arnold Arons[29-30] and other sources. Sometimes the problems were textbook problems or modified textbook problems. The problems were chosen to be on content that had already been covered by all of the instructors teaching the course. The problems were chosen to cover concepts or skills that students often struggle with, even after instruction.

The students would work through the problems in groups, working on whiteboards, with the TA circulating, asking students questions or answering questions from students. After students had had a significant amount of time to work on the problem, the TAs checked on students' understanding in different ways. Some TAs worked with groups individually, checking on their results both as they worked and as they finished, asking them to explain their results and asking further questions, as needed. Others called the class together and had groups present at the board and had a class discussion about the problems.

If there was time after the problem(s) for that week had been finished, the TAs entertained questions on homework or other questions students might have. The bi-weekly quizzes were also administered during the recitation sessions.

3. **TA training**

The TAs were trained and directed by the lab coordinator(s). They were taught to guide the students through questioning, not "telling" answers, but helping students to think through the questions themselves. They were taught how to help each group and to make sure everyone in the group contributed and was responsible for their own understanding. They were also taught how to guide whole class sessions, having groups or students present at the board and then lead class discussions. The teaching methods were modeled in their own TA training by the lab coordinator(s). The use of these interactive-engagement methods was expected of them both in the recitations and in the Exploration parts of the laboratories.

## B. The faculty

The majority of physics faculty members teach traditionally in a lecture-style format. Very few use PER-informed pedagogy or IE techniques. They focus primarily on the lecture and leave the recitations and laboratories to the TAs and lab coordinator(s). Although the labs and recitations were part of the course, the labs and recitations together were sometimes allotted as little as 10% of the grade. The lower allotments of the percentage of the grade for lab and recitation together were primarily in the calculus-based classes. However, some of the instructors allotted 20%-30% of the grade in those classes for lab and recitation (together). In the algebra-based classes, a higher percentage of the grade was allotted to the labs and recitation sections, with 20% the most common, although they ranged from 15% - 25%.

A few instructors interacted with the TAs in lab and recitations, contributing to the training of the TAs, the choice of materials and content to be covered in recitations and the pedagogy to be used in lab and recitation. Most of the instructors who actively participated in the TA training, were instructors who used PER-informed materials and instructional techniques in the lecture.

The instructors labeled by PER in this paper used PER-informed materials and teaching methods in the lecture.

## C. The students

## 1. Calculus-based Courses

## a. Large lecture sections of the calculus-based course

The calculus-based course consists primarily of engineering and computer science majors. The number of students registered for CBI, the first course in the sequence, each semester, is usually around 500, split among three lecture instructors. The number of students in the second course in the sequence, CBII, is around 400, split among two or three instructors. The instruction is primarily traditional lecture, with one one-hour recitation section and one two-hour lab, as described above. The labs and recitations are common among the three instructors each semester. Students from each of the lecture instructors are mixed in the labs and recitations.

**b. Honors section**

There is one honors section of the calculus-based class that is taken by students in the TTU Honors College and by some of the physics majors. It is usually a small class, consisting of 10 – 24 students. Sometimes the honors students take the same laboratories as the large lecture sections and sometimes they do not, depending on the instructor. We have listed the scores that we have for honors students who did take the same laboratories as the students in the large lecture sections with traditional lecture instruction. We also list the data for one honors section that worked through locally written PER-informed labs based on Workshop Physics[19] and other PER materials combined with PER lecture instruction, separate from the other sections. That course had an integrated lab/lecture format. The number of students in these sections is small and we hesitate to draw significant conclusions from the data because of the small number of students assessed. The results from the honors sections are included for completeness.

**2. Algebra-based Courses**

**a. Large lecture sections of the algebra-based course**

The algebra-based class consists mostly of pre-health science majors, including pre-medical, pre-dental, pre-physical therapy, etc. The number of students registered each semester in the first course in the sequence is usually around 250-300 and has been around 100-150 in the second course in the sequence in recent semesters[31]. Except for the inquiry-based section of the course, the students are divided into two lecture sections taught by two lecture instructors each semester. The instruction is primarily traditional lecture, with one one-hour recitation section and one two-hour lab each week. The labs and recitations are common among the three instructors. Students from each of the lecture instructors are mixed in the labs and recitations.

**b. Inquiry-based, laboratory-based section**

As described in the Introduction, an inquiry-based, laboratory-based section of the algebra-based course was developed with National Science Foundation (NSF) support[15-18] starting in 2001. The course was developed specifically for health science majors in the introductory algebra-based physics course. It is taught in a Workshop-Physics style[19] environment and is an inquiry-based course in the manner of <u>Physics by Inquiry</u>,[20] developed by the Physics Education Group at the University of Washington.

The curriculum was designed to be taught in a laboratory-based environment with no lecture and no text; however, a text can be used. Students work through the units in groups, learning about the world around them through experimentation, learning to develop both quantitative and qualitative models based on their observations and inferences. The materials consist of the laboratory units, pretests, readings and exercises. There are also homework sets, exams and quizzes. The students sign up on a first-come, first-serve basis.

The FCI, MBT, BEMA and CSEM were also administered to these students every semester, starting in Spring 2010.

## III. ADMINISTRATION OF ASSESSMENT INSTRUMENTS

The conceptual inventories were administered as pre- and post-tests in the recitations over the course of this study.  They were administered as a pre-test at the beginning of the semester and as a post-test at the end of the semester. Students were allotted 45 min. to take the assessments. The FCI and BEMA were administered as a pre- and a post-tests every semester starting in Fall 2009, except one semester (Spring 2011) when BEMA was administered as a post-test only.  We also have FCI data from semesters prior to Fall 2009 for select classes. The MBT was administered online as a pre- and a post-test in Spring 2010 and as a post-test only in Fall 2010 and Spring 2011. The CSEM was administered as a pre- and a post-test in Spring 2010 and as a post-test only in Fall 2010 and Spring 2011.

Taking the assessment counted as part of the students' laboratory or recitation participation grade. Depending on the class, one to three points were deducted from their participation grade if they did not take the assessment. In addition, starting in Fall 2010, up to five points toward their laboratory or recitation grade were awarded to students, based on their performance on the assessment. The number of points was determined by the percentage correct on the post-test. The students were not told their score and the assessment was not discussed with the students. They were simply told the number of extra points they received, if they asked. For most students this was three points or less. This constituted not more than 1% of their total course grade. We report any difference in the students' scores that could be accounted for by the year the assessment was taken. We present some of the ABI and CBI results chronologically in the next sections to address this concern.

The online administration of MBT and CSEM was terminated after one semester, Spring 2010, due to concerns about the efficacy of online testing, as well as concerns about too much assessment in the introductory courses.

**IV. RESULTS**

We report the results for each of the four conceptual inventories. We report pre-test, post-test and gain scores in Tables 1-4 and graph relevant post-test and normalized gain scores in Figures 1-14. The normalized gain,[24] g, was calculated using the equation

$$g = (\text{postscore} - \text{prescore}) / (100 - \text{prescore}).$$

The error bars represent the standard error of the mean.

**A. Force Concept Inventory**

The results for the normalized gain and the post-test scores for the FCI for both the algebra-based and calculus-based classes are presented in Figures 1 through 4. The data is presented in tabular form in Table 1. We have combined all of the data by lab and teaching style and present the means and standard error for each lab and lecture teaching style. For the algebra-based course, data from Fall 2006 through Fall 2011 are included, but the FCI was not administered every semester until Fall 2009. For the calculus-based course, the data includes Fall 2008 through Spring 2012, except for Spring 2009. We report both the normalized gain, pre- and post-test scores only for those students who took both the pre and post-test. The percent of total grade allotted to the combined lab/recitation part of the course is also given. In addition, we plot the data for the large lecture classes grouped chronologically in Figures 5 and 6, so that any differences that may be due to the small amount of credit given starting in Fall 2010 can be observed.

The data in the graphs is labeled by: *laboratory/teaching style (N = number of students) percentage total grade allotted to laboratories plus recitation (and chronological information in Figures 2 and 5)*. Lab styles are labeled by traditional (T), Real Time Physics (RT), combination RT and T (RT-T), developed at the University of Illinois (IL), and locally written PER-informed (PER). The lecture teaching styles are labeled by traditional lecture (TL), PER-informed lecture (PERL) and Inquiry-based instruction (INQ). Honors sections are labeled with an *H*.

The FCI data from traditionally taught labs (T) was collected before the recitation sections were introduced. FCI data from all other lab styles was taken after the recitations were implemented.

**1. Algebra-based FCI**

For the algebra-based course, we have FCI data with traditional (T) and PER-informed Real Time Physics labs (RT) with both PER-informed (PERL) and traditional (TL) lecture teaching styles. This gives us information on the effectiveness of the RT labs and recitation compared to T labs and PERL vs. TL lecture styles.

For comparison with other scores across the country, we refer to a large survey paper by Hake.[24] In that paper, it is reported that most students taught in a traditional lecture format have normalized gain scores of about 0.15 and students taught in an IE format generally have scores in the 0.30 – 0.60 range, known as the IE region on a "Hake" plot. Most PER-informed materials, such as those listed in the PER User's Guide,[32] employ IE methods and are also designed to address many of the alternative conceptions found on the FCI, so it is also expected that the use of PER-informed materials will result in an increased normalized gain.

The distributions were determined to be normal based on histograms of the data and we used a Student's T-test to determine if the data were significantly different. All of the results in the Figures and Tables are significantly different for comparison of data with non-overlapping error bars at the $p < 0.05$ level by a Student's T-test for all of the algebra-based scores.

## a. Effect of RT labs

The traditional (T) labs have a very low normalized gain with TL lecture instruction, 0.09 +/- 0.02. When RT labs and recitations were introduced, also with TL instruction, the normalized gain is still not particularly high, but it has increased by 100% above the T labs to 0.18 +/- 0.01. This effect is due to the RT labs plus recitation, in a situation where the lecture instructors did not pay much attention to the labs and recitations.

If we compare T labs to RT labs with PERL lecture instruction, the situation is similar. The normalized gain with T labs and PERL instruction is 0.22 +/-  0.02. When T labs are replaced by RT labs plus recitations, the gain increases to 0.36 +/- 0.02, a 64% increase. With both RT labs and PERL instruction, normalized gain is in the interactive engagement region on a "Hake" plot.

If the data are examined chronologically (Fig. 2), allowing a distinction of data taken before a small amount of credit was awarded based on post-test scores, we do see a significant difference for the RT/TL data, but not for the RT/PERL data. The RT/TL data taken before Fall 2010 had a normalized gain of 0.17 +/- 0.01 and after Fall 2010, it had a normalized gain of 0.20 +/- 0.01. It is possible that this difference is due to the credit given. However, both scores are still much lower than the normalized gain score for RT/PERL instruction and we do not see a significant difference in the RT/PERL scores before and after Fall 2010.

## b. Effect of PER instruction

We can also examine the effect of PER and TL instruction for different lab styles. For T labs, the normalized gain increases by 144% percent with PERL instead of TL instruction in the lecture, but still neither of the scores are very high, at 0.09 +/- 0.02 and 0.22 +/-  0.02, for TL and PERL, respectively. With the RT labs, a change from TL to PERL instruction results in a 100% increase in normalized gain, from 0.18 +/- 0.01 to 0.36 +/- 0.02, and, again, the score is now in the interactive engagement region on a "Hake" plot.

## c. Combined Effect of RT and IE and Comparison to Inquiry

It is only with the use of both RT labs and recitations and PERL lecture instruction that the FCI normalized gain is above 0.30, and into the interactive-engagement region. The mean FCI normalized gain for the Inquiry-based (INQ) class is 0.34 +/- 0.02 and not significantly different from that for the large lecture sections when both RT labs and PERL instruction are implemented. Both the INQ class and the large lecture sections with PERL instruction in the lecture and RT labs and recitation sections, demonstrate the effectiveness of research-based instruction.

The post-test scores are above 50% when both RT and PERL are introduced and in the INQ class. It is more common to compare normalized gain scores to others across the country, because FCI pretest scores can vary significantly. However, we present posttest scores also. Our local pretest scores were not significantly different from each other in any of the classes.

An important point here is that the FCI scores, while increased 100% above the T/TL scores when RT labs and recitations were introduced without changing the lecture instruction (RT/TL ), are still closer to traditional scores than to the IE region. This is important information for universities who try to implement PER-informed techniques in the labs and recitations only, without faculty concurrence to change instructional methods in the lecture also. The gains are not as large as they would be if faculty would overcome their hesitancy and resistance to the adoption of PER-informed techniques.

## 2. Calculus-based FCI

The data from the calculus-based classes is more complex. The transition to PER-informed labs in the calculus-based classes was hindered and very much opposed. The first introduction of PER-informed labs was part RT and part traditional labs (RT-T). RT labs only were run for only one semester before the introduction of labs developed at the University of Illinois (IL). So there are four lab styles, T, RT-T, RT and IL that have been used in the large lecture classes. Each of the lab styles has been run with both PERL and TL teaching techniques.

In addition, we report scores for the honors/majors sections when they took the same labs as the large lecture sections and for one honors section taught PER/PERL, separate from the other sections.

We also indicate the percent of the total grade allotted for lab plus recitation because it varied somewhat, from 12.5% to 20%. However, we do not think we can draw any conclusions about the effect of the percentage grade allotted for lab and recitation, as we can in the second semester BEMA data presented in Section IV.C.2. We graph the data chronologically in Figure 5 to examine if there is any change that could be due to a small amount of credit awarded for correct answers on the post-test. There is no significant difference between pre-Fall 2010 and post-Fall 2010 data for the PERL instructional method, independent of the lab method used. There are differences by lab method when the lecture is traditional (TL). We believe these differences are more likely due to the lab method as opposed to the small amount of credit awarded post-Fall 2010, but cannot make a definitive statement based on the pre-Fall 2010 and post-Fall 2010 data.

The large lecture sections barely achieve normalized gains of 0.30 with PERL instruction (any type of lab) or RT labs with TL instruction. The PERL data and the RT/TL instruction data are not significantly different. The T, IL and RT-T labs with TL instruction are not significantly different from each other by a Student's T-test at the $p < 0.05$ level, but they are each significantly different from the RT/TL labs at the $p < 0.0001$ level.

The traditional (T) labs were taught without recitation. The T/TL and T/PERL comparison then gives us information on the implementation of PER-informed instruction in the lecture as the only change. There is a 93% increase in the normalized gain when PER-informed instruction is implemented compared to traditional instruction.

While we don't have information on the RT, RT-T and IL labs without recitation, the recitation was implemented the same way for all of the lab styles. For TL instruction, we observe a difference in the normalized gain based on the lab style, with the RT labs resulting in much higher normalized gain than the RT-T and IL labs. This difference is a result of the lab style only. We do not see this difference when the lecture style is PERL. This again indicates the importance of implementing PERL instruction in the lecture and not just in the laboratories and recitations.

Only the honors/majors sections achieve gains well above 0.3, in the IE region on a "Hake" plot, with the honors section taught by integrated PER/PERL methods achieving a normalized gain of 0.5. While there are many differences between the honors/majors section and the large lecture classes, the comparison of the different lecture styles among the honors classes (even though the number of students is small), again indicates the normalized gain is higher for the class with PER-informed instruction in both lecture and lab, as we see in the large lecture sections and in the algebra-based courses.

All of the distributions were determined to be normal by examining histograms of the data, even for the smaller classes. The labs with TL lecture instruction were not significantly different from each other, except for the RT/TL labs, which were significantly different from the other labs with TL lecture instruction by a Student's T-test at the $p < 0.0001$ level.

**B. Mechanics Baseline Test**

The MBT was also administered as part of this study. However, it was not administered every semester and was sometimes administered only as a post-test or only online. It was administered online in Spring 2010 and only as a posttest in Fall 2010 and Spring 2011.

The posttest results are presented in Figures 7 and 8 and Table 2, with the online results separated out. The online results were lower for the TL lecture classes, but not for the PERL lecture classes. This was a semester when a small amount of credit was not given for correct answers on the post-test and that could account for the difference in the TL lecture sections. However, we do not see the effect in the PERL data in the calculus-based classes and there may be other factors that contribute to this effect also.

There is not as much published data for comparison to scores at other universities as there is for the FCI. Students have been reported to score on the order of 15 percentage points lower on the MBT than the FCI and it has been considered to be a harder test and have more problem solving in it, as it requires some math skills and some critical thinking skills.[11]

**1. Algebra-based MBT**

The online scores differ on the order of 10-15 percentage points between in-class and online testing for the large lecture sections. This could also be due to the difference in offering a small amount of credit for correct answers. However, the scores for the INQ course are not significantly different online and in-class. All of the distributions were determined to be normal by examining histograms of the data, even for the smaller classes. The RT/PERL in-class scores are significantly different from the RT/TL scores at the $p < 0.003$ level and from the INQ scores at the $p < 0.002$ level on a Student's T-test. For the in-class scores, RT/PERL classes have scores in the high forties, RT/TL classes in the low forties and the Inquiry class is in between. There is not much comparison data for the algebra-based course. Scores as high as the high sixties have been observed in algebra-based classes using Peer Instruction (PI)[33] at Harvard University.[34]

**2. Calculus-based MBT**

As with the algebra-based scores, the RT-T/TL students who took the assessment online scored about ten points lower than students in any of the classes who took the assessment in-class. All of the distributions were determined to be normal by examining histograms of the data, even for the smaller classes. The online RT-T/TL are significantly different from all of the other data at the $p < 0.003$ level on a Student's T-test. The only other significant difference is between the in-class IL/TL and in-class RT/TL data at the $p < 0.004$ level. The scores of students who took the assessment in class are for the most part in the high forties. This is at the low end of scores published nationally,[11,34] which range from the forties to the high seventies. Classes taught traditionally fall on the lower end of that scale and classes taught by PI at Harvard University fall at the high end of the scale. The TTU scores are seven or eight points below the TTU FCI scores.

## C. Brief Electricity and Magnetism Assessment

BEMA was administered as a pre- and a post-test every semester from Fall 2009 through Spring 2012 in the calculus-based course, except for Spring 2011, when it was administered as a posttest only. In the algebra-based course, it was administered as a pre- and a post-test every semester from Fall 2009 through Fall 2011, except for Spring 2011, when it was administered as a post-test only.

In the Spring of 2011, in the large lecture sections in the algebra-based course, we had evidence of cheating on the BEMA post-test in the form of a TA talking with students during the assessment and some students with identical high grades, inconsistent with the rest of their work. We removed all of the data from that TA's sections and obvious cheating from other sections. However, we do not know how widespread the cheating was and if we have removed all of it, so we have chosen not to present the data from Spring 2011 for the algebra-based course.

The results for the normalized gain and the post-test scores for BEMA for both the algebra-based and calculus-based classes are presented in Figures 9 through 12. While it is very common to report only the BEMA post-test, as the pre-test is usually around 22% for all classes, we present both the normalized gain and the post-test scores. The data is presented in tabular form in Table 3, also including the pretest data. We have combined all of the data by lab and teaching style and present the means and standard error. We report both the normalized gain and pre- and post-test scores only for those students who took both the pre and post-test. The percent of total grade allotted to the combined lab/recitation part of the course is also given. This is particularly relevant in the calculus-based classes.

## 1. Algebra-based BEMA

In the algebra-based classes, we had already been using RT labs with some locally written labs when we began the assessment using BEMA. We do not have a comparison to the algebra-based courses taught with traditional labs. The results are not particularly high, with the highest gain at 0.17 and the highest post-test at 35.7%. The locally written, PER-informed labs (PER) have a higher gain and post-test than the RT labs with TL instruction, and even greater with PERL instruction. We do not have the RT labs with PERL instruction for comparison. The RT/TL data was taken before Fall 2010 when some small credit was given for correct answers on the post-test, so we cannot say that differences of those data are not due to that fact. However, we still see the very significant difference between PERL and TL lecture instruction. The distributions were determined to be normal based on histograms of the data and we used a Student's T-test to determine if the data were significantly different. All of the results in the Figures and Tables for comparison of data with non-overlapping error bars are significantly different at the $p < 0.01$ level by a Student's T-test. The INQ and PER/PERL scores are not significantly different.

There are not a lot of comparison scores in the literature for the algebra-based course. Most of the research using BEMA has been done with calculus-based classes. Typical post-test scores reported for calculus-based students are in the 40-50% range for traditionally taught students and around 60% for students taught non-traditionally with research-based materials.[35-37] We have found one algebra-based score of 0.38 gain and 51% post-test score posted on a Physics Teacher Education Coalition (PhysTEC) website.[38] While our scores are not particularly high, they have increased in the large section algebra-based classes, as we introduced locally developed PER-informed laboratories and used PER-informed instruction in the lecture.

## 2. Calculus-based BEMA

In the calculus-based classes, we have data with traditionally (T) taught labs and locally developed PER-informed laboratories (PER). The T/TL data was taken before Fall 2010 when some small credit was given for correct answers on the post-test, so we cannot say that differences of those data are not due to that fact. The distributions were determined to be normal based on histograms of the data and we used a Student's T-test to determine if the data were significantly different. All of the results in the Figures and Tables for comparison of data with non-overlapping error bars are significantly different at the $p < 0.005$ level by a Student's T-test. Since we have a record of the percentage the lab and recitation together counted towards the total course grade, it is interesting to examine the difference between lab/recitation counting as 10% or 30% of the grade. The BEMA scores with T labs were not significantly different from the PER labs when the labs plus recitation were allotted 10% of the total grade with TL lecture instruction. If the percentage grade allotted to the labs/recitation was raised to 30%, the PER BEMA scores were higher and significantly different (at the $p < 0.005$ level) from the T scores with 10% allotted to labs/recitation and TL instruction. With PER, instead of TL lecture instruction, the gain rises to close to 0.20 and the post-test to close to 40. These scores are consistent with scores reported for traditionally taught students at other universities across the country.[35-37] The honors physics class has a gain of 0.25 and a post–test of 43%.

## D. Conceptual Survey of Electricity and Magnetism

The CSEM was administered as a pre- and a post-test in Spring 2010 and Spring 2011 and as a post-test only in Fall 2010. The Spring 2010 assessment was administered online and we do not know if that was the reason for lower scores. We have chosen not to present the Spring 2011 data, due to the issues with cheating, as discussed with the Spring 2011 BEMA scores. We also do not have CSEM scores with PER instruction and we do not have CSEM scores with 30% of the total grade allotted to labs and recitation. As a result, our CSEM scores are not comprehensive. There has been research demonstrating statistically indistinguishable gains on BEMA and CSEM,[39] so it is plausible that the CSEM results in categories not assessed would be similar, but we do not know that for sure. In addition, the algebra-based RT/TL and the calculus-based T/TL were administered before a small amount of credit was given for correct answers on the post-test.

We present the normalized gain and post-test scores in Figures 13 and 14, including all of the available algebra-based and calculus-based scores on one plot for the normalized gain and another for the post-test scores. The distributions were determined to be normal based on histograms of the data and we used a Student's T-test to determine if the data were significantly different. All of the results in the Figures and Tables for comparison of data with non-overlapping error bars are significantly different from each other at the $p < 0.005$ level by a Student's T-test.

In this case, the online results were lower (and did not include a small amount of credit for correct answers) and the other results, calculus-based or algebra-based, were not significantly different from each other, with scores close to a gain of 0.2 and a post-test of 40%. The honors students scored higher. Common post-test scores for students nationally are in the 40's for both the algebra-based and calculus-based classes, although higher and lower scores, 30's for algebra-based and 50's for calculus-based, have been reported.[40-41]

## V. DISCUSSION

This project is important because it provides data on the introduction and implementation of PER-informed materials into the labs and recitation sections at a large university. The faculty were not unified and in agreement on the implementation of PER-informed materials and most faculty continued to teach traditionally in the lecture portion of the course, leaving the implementation of new materials in the laboratory and recitation to the TAs and the lab coordinator(s).  As there were a few course instructors who did use PER-informed materials in the lecture and had significant interactions with the lab coordinator and TAs, the study reflects the impact of changes when PER-informed materials are introduced in the laboratory and recitation, with and without changes in the lecture part of the course.

## A. Major findings

The data from the FCI in the algebra-based course clearly demonstrate an increase in conceptual understanding, as measured by the FCI, due to the implementation of PER-informed materials and instruction in the laboratories and recitations and due to changes from TL to PERL teaching methods in the lecture. Changes in the laboratories and recitations only or changes in the lecture only, both significantly increase normalized gains above T/TL instruction, with the change to PERL instruction in lecture only having a somewhat greater effect than changes in the labs only. Only with the implementation of PER materials in both the lecture and the laboratory and recitations, are normalized gains above 0.30, in the IE region on a Hake[24] plot observed in the large lecture classes.

Results from the calculus-based classes are similar; the normalized gain increases with either implementation of PER-informed materials and instruction in the laboratories and recitations or changes from TL to PERL teaching methods in the lecture. In this case, the completely PER-informed laboratories (RT) with traditional instruction increased the conceptual understanding as much as PERL instruction in the lectures with any kind of lab. However, unlike the algebra-based course, the combination RT/PERL was not significantly higher than either of those changes independently. The PER/PERL honors section, taught with integrated lab/lecture instruction and completely separate from the other sections, had the highest normalized gain on the FCI.

The FCI gains in the large lecture algebra-based classes with PER-informed instructional techniques in both lab/recitation and lecture were at least as high or higher than in the calculus-based courses (except for the honors sections). As the implementation of the interventions was similar in the two groups, it is worth further research into the reasons for the differences to see if other factors, such as goals, learning styles, expectations or motivations, play a role.

The data from BEMA indicate the same increase in normalized gain that we see with FCI, as PER-informed labs are added and with PERL instructional methods both in the lecture and the laboratories. However, neither the algebra-based nor the calculus-based classes achieve more than 0.20 in normalized gain or above 40% on the post-test, except for the honors sections. The scores, for the most part, are significantly below other published scores from universities introducing new curricula, such as instruction from the Matter and Interactions[42] curricular materials.[35-37]

We did, however, in the BEMA study, see the impact of the percentage of the total grade allotted to the laboratories and recitation sections. With TL instruction in the lecture, the changes to the labs and recitation sections made a significant difference, if the labs and recitations accounted for 30%, as opposed to 10%, of the total course grade.

The MBT data is fairly flat and the CSEM data all falls within the ranges seen across the country.

**B. On the use of conceptual inventories and further assessment**

In this paper, we have presented the results of conceptual inventories as indicators of the effectiveness of changes made in the curricula and instructional techniques in the laboratories and recitations separately and together with changes made in the lecture part of the course. The conceptual inventories are designed to assess conceptual understanding and it is reasonable to expect that PER-informed changes to instruction would result in increased gain in conceptual understanding. However, one would hope that PER-informed changes would result in changes in many skills, from lab skills to computational skills, to problem solving and critical thinking skills, not just changes in conceptual understanding. One might argue that conceptual inventories, while giving significant information on changes in conceptual understanding, are not the best instruments to assess some of these other skills.

Based on conceptual inventory scores, the inquiry-based class and the large lecture classes with PER-informed labs and PERL lecture instruction performed at the same level. Based only on conceptual inventory data, the two methods of instruction lead to identical results. But is this the whole story? How would the two methods of instruction compare if other assessment instruments were used to assess other skills?

Too often, conceptual inventory results are presented as if they are comprehensive results, the main factor in determining whether instruction has been effective or not. We believe that this use of conceptual inventory results is unfortunate and that different and more comprehensive assessment instruments need to be developed by the PER community. We support researchers developing assessments that go beyond conceptual inventories, assessing problem solving, lab skills and other important aspects of instruction.[43-44]

In our project, we also administered a series of free-response pre- and post-tests in the labs and recitations over four semesters. The questions required written explanations, covered lab, recitation and lecture concepts and also assessed lab skills and problem solving. They were administered bi-weekly and were not comprehensive. They did, however, give us snapshots of students' understanding and abilities throughout the course. This research gave us different information on the students' abilities when different instructional methods were used. The results of that research are presented in other papers.[23]

## VI. CONCLUSIONS

We conclude that when PER-informed materials are introduced through the labs and recitations, independent of the lecture style, in a large university setting, there is an increase in students' conceptual understanding, as measured by PER developed conceptual inventories. There is also an increase in the results on conceptual inventories, if PER-informed instruction is used in the lecture. The highest normalized gains were achieved by the combination of PER-informed lectures and laboratories in large class settings and by a hands-on, laboratory-based, inquiry-based course and a PER/PERL taught honors section, both in small class settings. We hope that these results will motivate change at our own and similar institutions and be informative to PER researchers studying barriers to change and those working on assessments beyond conceptual inventories.

## VII. ACKNOWLWDGEMENTS

We thank Kelvin Cheng and Amy Pietan for their work on the statistical analysis that contributed to this paper. We also thank the National Institutes of Health (NIH) for their support of this project, for the funding of NIH Challenge grant #1RC1GM090897-02. Any opinions, findings and conclusions or recommendations expressed are those of the authors and do not necessarily reflect the views of the NIH.

**Appendix I**

**LABORATORY 7**
**MAGNETISM I: MAGNETIC FIELDS**

**Objectives**

- to be able to represent a magnetic field by appropriately drawn magnetic field lines

- to observe that a current gives rise to a magnetic field

- to be able to determine the direction of the magnetic field due to a current-carrying wire

- to be able to recognize and discuss the superposition of magnetic fields

- to be able to determine quantitatively and discuss qualitatively the dependence of the magnitude of the magnetic field with distance from a current-carrying wire

- to be able to measure the strength of a magnetic field as a function of distance from a current-carrying wire

- to be able to discuss qualitatively and determine quantitatively the magnetic field near the center of a solenoid

**Overview:** In this laboratory, we will observe that a current gives rise to a magnetic field and determine the direction and magnitude of the magnetic field by observation and measurement.

**Equipment:**
      1 permanent magnet
      6 small compasses
      10-12 square stack magnets

**Exploration 1 Magnetic fields of magnets**

**Exploration 1.1 The direction of the magnetic field of a magnet**

    **a.** Consider a bar magnet. Draw the magnetic field lines for a bar magnet in the diagram below.

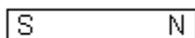

**b.** Consider a small magnet, like one of the stack magnets at your table. Where are the poles of the stack magnet? Test for the poles with other magnets or compasses. Label the poles of the magnet and then draw the field lines for one of the small magnets at your table.

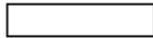

Check your labeling of the poles with your TA.

**c.** Now consider a stack of magnets, as in the diagram below. Label the north and south poles and draw the field lines. Explain why you drew the field lines the way you did.

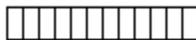

**d.** Remove the left half of the stack as in the diagram below.

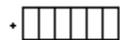

Draw the direction of the magnetic field at the point to the left of the magnet in the diagram.

**e.** Replace the left half and remove the right half as in the diagram below.

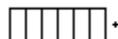

Draw the direction of the magnetic field at the point to the right of the magnet in the diagram.

**f.** By superposition, when both halves are in place, the magnetic field points in the direction of the sum of all magnetic fields at that point. Which direction does the magnetic field point inside a magnet? Explain.

## Exploration 1.2

**a.** Draw the magnetic field of a horseshoe magnet in the diagram below.

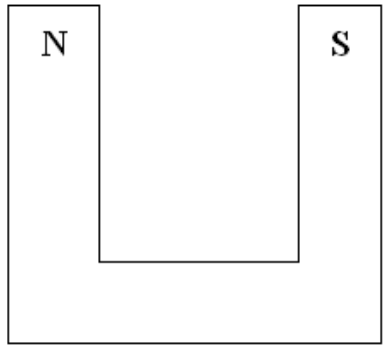

Discuss your drawing with a TA.

**Equipment:**
    1 wire stand
    6-9 small compasses
    1 power supply
    1 3-dimensional compass

### Exploration 2: The magnetic field of a current-carrying wire

### Exploration 2.1 The Earth's magnetic field

Consider the set up at your table that consists of a wire, a stand and a power supply, as in the diagram below.  The wire passes through a hole in the wood that is supported by the stand.  The power supply is connected to the wire and a current can be sent through the circuit, when the power supply is turned on. DO NOT turn on the power supply until instructed to do so. With the power supply turned off, place about 6 compasses on the wood in a small circle around the wire.

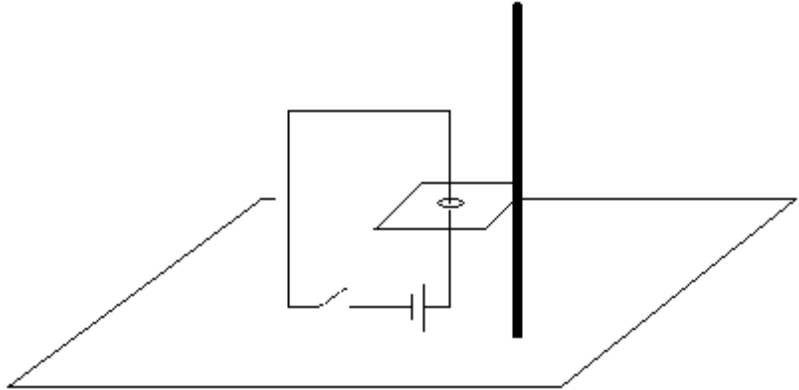

**a.** With the power supply turned off, which direction do the compasses point? Do they all point in the same direction?

**b.** A compass is a small magnet. It will align itself with a magnetic field. What produces the net magnetic field with which it is aligned? Do you think there is more than one magnetic field in the region of the compasses? Do you think there is a dominant magnetic field in the region of the compasses? Explain.

**c.** If the compasses were not constrained to rotate only in the plane, but could rotate in three dimensions, do you think they would point in the same direction as they do now? Predict the direction a three-dimensional compass would point.

**d.** Use the three-dimensional compass at your table to test your prediction.

## Exploration 2.2 Direction of the magnetic field of a current-carrying wire

**a.** This is a prediction. Do not carry it out yet. Do not turn on the power supply. Predict what would happen to the compasses, if the power supply were turned on and there was a current in the wire.

**b.** Turn on the power supply and turn the current up to about 0.75A. DO NOT turn the current up over 1.0 A. (There is very little resistance in the wire and you will burn it up, if you turn the current on too high or leave it on for a long period of time.) Observe what happens to the compasses. Do not the leave the current on for a long period of time, just long enough to observe the compasses, then turn the current off. Record what happened below. Is there a pattern to the direction of the compasses?

**c.** Connect the power supply so that the current travels the other direction through the wire. Turn the current on again. Observe what happens, then turn the current off. How is the pattern of the compasses the same or different as in part **b**?

**d.** Based on your knowledge of compasses as small magnets, is there a magnetic field near the wire

(i) when current is not flowing through the wire?

(ii) when current is flowing through the wire?

Explain your reasoning.

**e.** Is there more than one magnetic field in the region of the compasses when the current is flowing through the wire? Explain.

**f.** If there is more than one field near the current-carrying wire, discuss the strength of the fields, based on your observations.

**g.** Discuss the approximate direction of the magnetic field of a current-carrying wire, if the current is moving up the wire in the region near the compasses.

**h.** Discuss the approximate direction of the magnetic field of a current-carrying wire, if the current is moving down the wire in the region near the compasses.

## Exploration 2.3 Magnitude of the magnetic field

**a.** Place the compasses on the wood in a line away from the wire, each at a different distance from the wire.

**b.** Does this give you any information on the magnitude of the magnetic field? Where is it weak and where is it strong? Explain.

## Exploration 2.4 The field of a wire loop

**a.** Suppose a single current-carrying wire was bent into a loop as in the diagram below. Draw the magnetic field lines for the area around the loop (inside and outside the loop) in the diagram below.

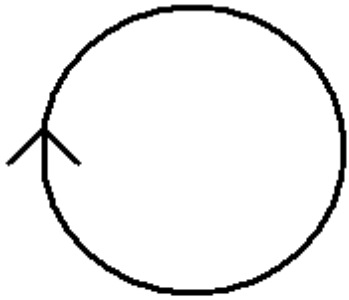

wire with current I flowing clockwise

**b.** Consider a sideview of the same loop. Draw the magnetic field lines. In the picture below.

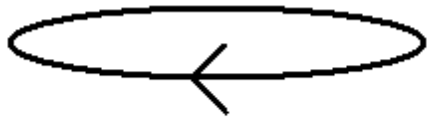

side view of the top loop

**c.** Compare the field lines for a closed loop to that of the magnets above. Are there any similarities with the field lines of any of the magnets?

**Exploration 2.5 The magnetic field of a solenoid**

**a.** Draw the field lines for a solenoid, a series of closed loops, in the diagram below.

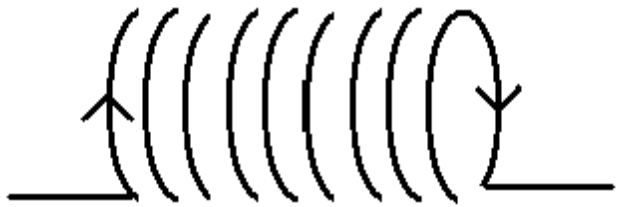

**b.** Compare the field lines of a solenoid to that of the magnets above. Is there a similarity to any of the magnets? Which ones?

**Equipment:**
>1 wire mounted vertically on a stand
>1 power supply
>1 magnetic field sensor
>1 computer interface
>1 computer software

**Investigation 1 The magnetic field of a current carrying wire**

Theoretically, the value of the magnetic field of a current carrying wire is

$$B = \frac{\mu_0\, I}{2\pi r}$$

Where I is the current, r is the distance from the wire and $\mu_0$ is a constant. $\mu_0 = 4\pi \times 10^{-7}$ nm$^2$/C$^2$. We are going to measure the magnitude of the field of a current-carrying wire at different distances from the wire using a magnetic field sensor.

**Investigation 1.1 The magnitude of the magnetic field of a current carrying wire**

   **a.** If you were to plot the magnitude of the magnetic field vs. the distance from the wire, what kind of graph you would expect? Sketch the shape of the graph in the space below.

   **b.** Would the plot in **a.** be the best way to verify how the magnetic field depends on the distance from the wire? Is there any other way you could plot the data? Explain.

   The magnetic field sensor is mounted on a rotary motion sensor, so you can easily move it a small distance at a time. It should be connected to the computer interface and the computer  interface connected to the computer. Bring up Data Studio. Click on Create and Experiment. Click on one of the digital input channels on the left. Select Rotary Motion Sensor. Click on the second "Measurement" tab. Choose Position. Click on Analog Channel A. Choose Magnetic field Sensor. Select Magnetic Field Strength 100x. Select units Tesla.

   Align the Sensor. The Sensor should be radial along ruler on stand, offset 7mm from the wire when sensor casing is touching wire. See picture on board.

   **c.** Before turning the power supply on (no current through the wire), would you record a magnetic field, if you took data? Why or why not? Explain.

Test your answer, by hitting the Start to take data.

If there is a field, what is the magnitude of the magnetic field in mT? Does the magnitude make sense? Explain.

If you would like to record only the field of the current-carrying wire, you can zero the sensor when there is no current through the wire by pressing the "Tare" button on the sensor. Press the "Tare" button now.

**d.** Now turn on the current through the wire to about 0.75A. Take data by hitting "Start" and moving the sensor very slowly away from the wire.

**e.** View the data on a graph and bring up a table of the data by sliding the graph and table icons to the right of the screen.

Copy the data from the table to an Excel file.

Delete any early data with position zero.
Delete any data taken after the motion has stopped.

Shift the position data by 0.007m (to account for the initial offset of the probe).

The magnetic field should not have any negative values, since you zeroed the sensor. However, if you do have negative values, you will have to add a constant to all of the data, so that none of the values are negative.

Plot B vs. r and fit the data to a power law.

**f.** Plot B vs. (1/r) and fit the data to a straight line.

**g.** Find $\mu_0\,I/2\pi$ from each of your plots. You should use 25 times the current value you read on the power supply because the wire really consists of 25 wires. Show your calculations below.

**h.** Determine the value of $\mu_0/2\pi$ from each graph. Show your work below.

**Equipment:**
    1 solenoid
    1 power supply
    1 magnetic field sensor
    1 computer interface
    1 computer software

**Investigation 2 The magnetic field of a solenoid**

In this section, we will measure the magnetic field of a solenoid and use it to determine the number of turns of the solenoid.

**Investigation 2.1 The magnetic field of a solenoid**

    **a.** Discuss the strength of the magnetic field inside a solenoid. Is it constant?

    **b.** If you were to measure the magnetic field inside a solenoid, would it be uniform everywhere?

    **c.** Is the field inside the solenoid approximately uniform in some region?

    **d.** Test your predictions with the magnetic field sensor.

**Investigation 2.2 Measuring the magnetic field of a solenoid**

Theoretically, the magnetic field near the center of a solenoid is

$$B = \mu_0 n I$$

where I is the current and n is the number of turns per unit length.

We are going to determine the number of turns per unit length of the solenoid by measuring the magnetic field in the center of the solenoid.

    **a.** Use the magnetic field sensor to measure the field inside the solenoid at your table for 5-6 values of the current through the solenoid. Keep the current below 1.0A as much as possible. Record your data in the table below.

| Current (A) | Magnetic field (T) |
|---|---|
|  |  |
|  |  |
|  |  |
|  |  |
|  |  |
|  |  |
|  |  |

**a.** In Excel, plot your data and determine the *n*, the number of turns per unit length, of the solenoid. Describe your method for determining *n* and show your work clearly in the space below.

**Summary.** Summarize how you determined the magnitude and direction of the magnetic field due to a current-carrying wire in a few sentences in the space below.

### Laboratory 7 Homework
### Magnetic Fields

1) Two long wires lie parallel in the plane of the paper, with current in the direction indicated.

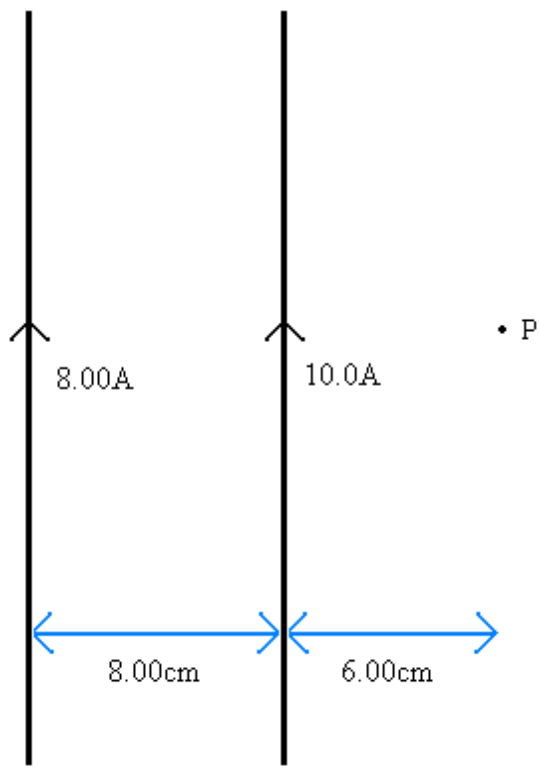

a) Calculate the magnitude and direction of the net magnetic field at Point P. Show your work and explain your calculations.

b) Calculate the magnitude and direction of the net magnetic field halfway between the two wires. Show your work and explain your calculations.

| Lab/Lecture Style | N | %Lab/Rec. | Pretest | S.E. | Posttest | S.E. | g | S.E. |
|---|---|---|---|---|---|---|---|---|
| Alg.-based | | | | | | | | |
| T/TL | 102 | 15% | 30.9 | 1.3 | 37.4 | 1.5 | 0.09 | 0.02 |
| RT/TL | 994 | 20% | 29.5 | 0.4 | 42.2 | 0.6 | 0.18 | 0.01 |
| T/PERL | 79 | 20% | 30.5 | 1.3 | 45.5 | 2.0 | 0.22 | 0.02 |
| RT/TL | 162 | 20% | 30.2 | 1.0 | 54.4 | 1.4 | 0.36 | 0.02 |
| INQ | 116 | N/A | 29.4 | 1.3 | 53.0 | 1.7 | 0.34 | 0.02 |
| Calc.-based | | | | | | | | |
| T/TL | 113 | 20% | 36.6 | 1.2 | 46.1 | 1.3 | 0.14 | 0.02 |
| IL/TL | 556 | 15/20% | 40 | 0.6 | 49.5 | 0.8 | 0.16 | 0.01 |
| RT-T/TL | 377 | 12.5% | 37.6 | 0.8 | 50.1 | 0.9 | 0.20 | 0.01 |
| RT/TL | 236 | 12.5/20% | 39.4 | 1.2 | 56.5 | 1.3 | 0.28 | 0.02 |
| T/PERL | 25 | 20% | 36.3 | 3.3 | 54.2 | 3.6 | 0.27 | 0.05 |
| IL/PERL | 146 | 15% | 39.5 | 1.2 | 56.8 | 1.6 | 0.30 | 0.02 |
| RT-T/PERL | 36 | 20% | 40.6 | 3.1 | 57.4 | 3.7 | 0.31 | 0.05 |
| RT/PERL | 122 | 12.5% | 39.3 | 1.5 | 55.25 | 1.8 | 0.26 | 0.03 |
| H. IL/TL | 17 | 20% | N/A | N/A | 64.1 | 3.8 | N/A | N/A |
| H. RT/TL | 17 | 12.5% | 55.1 | 4.4 | 74.8 | 2.9 | 0.40 | 0.06 |
| H. PER/PERL | 18 | 15% | 54.8 | 5.0 | 75.2 | 4.7 | 0.50 | 0.02 |

Table 1. Results for FCI for algebra-based and calculus-based courses by lab and teaching style. Lab styles are labeled by traditional (T), Real Time Physics (RT), combination RT and T (RT-T), Illinois (IL), and PER-informed (PER). The teaching styles are labeled by traditional (TL), PER-informed (PERL) and Inquiry-based (INQ). Honors sections are labeled with an *H*.

| Lab/Lecture Style | N | %Lab/Rec. | Posttest | S.E. |
|---|---|---|---|---|
| Alg.-based | | | | |
| Spr. 10 RT/TL | 198 | 20% | 34.3 | 1.0 |
| Spr. 10 INQ | 15 | N/A | 39.0 | 4.0 |
| RT/TL | 301 | 20% | 42 | 0.9 |
| INQ | 38 | N/A | 43 | 2.5 |
| RT/PERL | 71 | 20% | 48.1 | 1.7 |
| Calc.-based | | | | |
| Spr. 10 T-RT/TL | 303 | 12.5% | 37.4 | 2.4 |
| Spr. 10 T-RT/PERL | 36 | 20% | 48.4 | 2.7 |
| RT/TL | 236 | 12.5/20% | 48.6 | 0.9 |
| IL/TL | 121 | 15/20% | 43.2 | 1.2 |
| RT/PERL | 122 | 15% | 47.0 | 1.4 |
| IL/PERL | 146 | 15% | 47.0 | 1.3 |

Table 2. Results for MBT posttest for algebra-based and calculus-based courses by lab and teaching style. Lab styles are labeled by traditional (T), Real Time Physics (RT), combination RT and T (RT-T), Illinois (IL), and PER-informed (PER). The teaching styles are labeled by traditional (TL), PER-informed (PERL) and Inquiry-based (INQ). Honors sections are labeled with an *H*.

| Lab/Lecture Style | N | %Lab/Rec. | Pretest | S.E. | Posttest | S.E. | g | S.E. |
|---|---|---|---|---|---|---|---|---|
| Alg.-based | | | | | | | | |
| RT/TL | 192 | 20% | 23.6 | 0.5 | 28.9 | 0.7 | 0.07 | 0.01 |
| PER/TL | 64 | 20% | 22.7 | 0.9 | 31.7 | 1.4 | 0.11 | 0.02 |
| PER/PERL | 58 | 20% | 21.9 | 0.9 | 34.7 | 1.7 | 0.17 | 0.02 |
| INQ | 62 | N/A | 20.7 | 1.0 | 35.7 | 1.4 | 0.16 | 0.02 |
| Calc.-based | | | | | | | | |
| T/TL | 345 | 10% | 21.9 | 0.4 | 27.5 | 0.6 | 0.07 | 0.01 |
| PER/TL | 241 | 10% | 21.7 | 0.5 | 30.4 | 0.7 | 0.08 | 0.01 |
| PER.TL | 105 | 30% | 23.84 | 0.9 | 33.7 | 1.3 | 0.13 | 0.01 |
| PER/PERL | 200 | 25% | 24.3 | 0.6 | 39.0 | 1.0 | 0.19 | 0.01 |
| H PER/TL | 9 | 20% | 25.2 | 2.5 | 43.3 | 2.3 | 0.25 | 0.05 |

Table 3. Results for BEMA for algebra-based and calculus-based courses by lab and teaching style. Lab styles are labeled by traditional (T), Real Time Physics (RT), combination RT and T (RT-T), Illinois (IL), and PER-informed (PER). The teaching styles are labeled by traditional (TL), PER-informed (PERL) and Inquiry-based (INQ). Honors sections are labeled with an *H*.

| Lab/Lecture Style | N | %Lab/Rec. | Pretest | S.E. | Posttest | S.E. | g | S.E. |
|---|---|---|---|---|---|---|---|---|
| Alg.-based | | | | | | | | |
| RT/TL | 138 (147 post) | 20% | 24.2 | 0.8 | 32.2 | 1.2 | 0.10 | 0.01 |
| PER/PERL | 53 | 20 % | N/A | N/A | 38.9 | 2.4 | N/A | N/A |
| INQ | 38 (58 post) | N/A | 24.3 | 1.5 | 38.2 | 2.1 | 0.2 | 0.03 |
| Calc.-based | | | | | | | | |
| T/TL | 165 | 10% | 26.13 | 0.8 | 30.8 | 1.1 | 0.06 | 0.01 |
| PER/TL | 163 (404 post) | 10% | 25.6 | 0.7 | 39.2 | 0.8 | 0.18 | 0.01 |
| H PER/TL | 12 | 10% | 27.5 | 1.0 | 53.62 | 3.1 | 0.36 | 0.04 |

Table 4. Results for CSEM for algebra-based and calculus-based courses by lab and teaching style. Lab styles are labeled by traditional (T), Real Time Physics (RT), combination RT and T (RT-T), Illinois (IL), and PER-informed (PER). The teaching styles are labeled by traditional (TL), PER-informed (PERL) and Inquiry-based (INQ). Honors sections are labeled with an *H*.

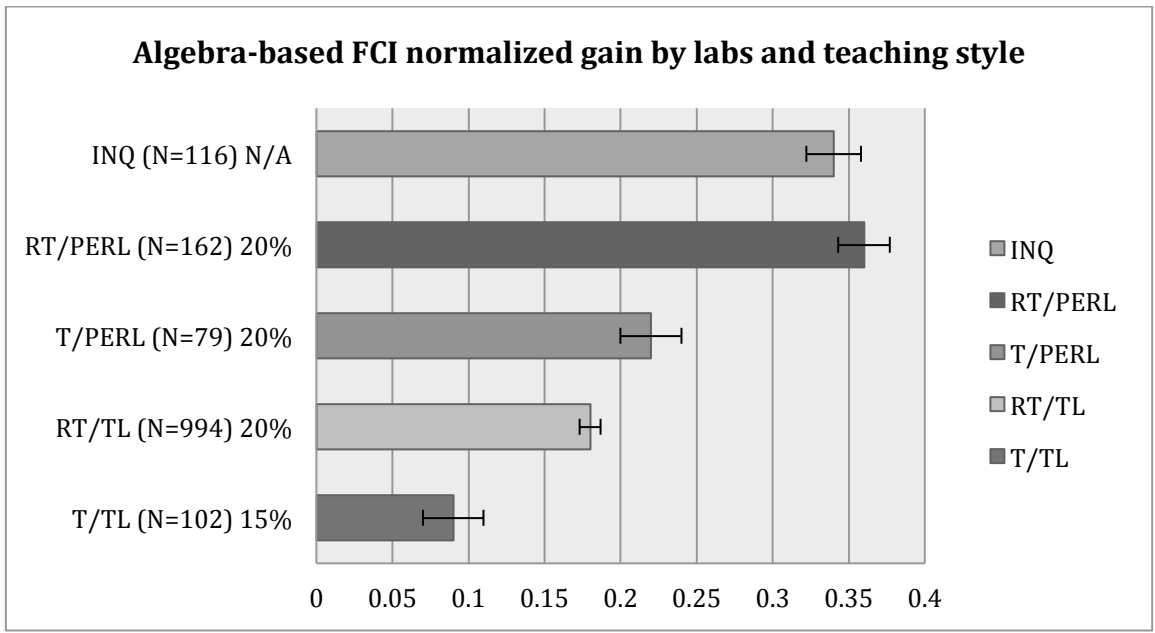

Figure 1. Algebra-based FCI gain by laboratory and teaching style. The data is listed by: *laboratory/teaching style (N = number of students) percentage total grade allotted to laboratories plus recitation.* Lab styles are labeled by traditional (T), Real Time Physics (RT), combination RT and T (RT-T), developed at the University of Illinois (IL)[26], and locally written PER-informed (PER). The lecture teaching styles are labeled by traditional lecture (TL), PER-informed lecture (PERL) and Inquiry-based instruction (INQ). Honors sections are labeled with an *H*.

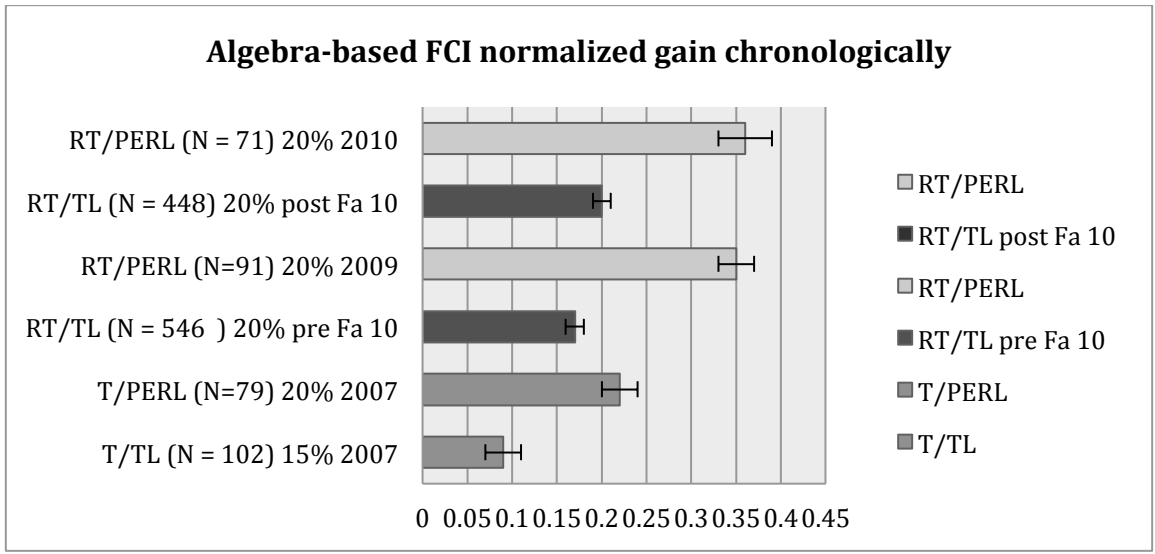

Figure 2. Algebra-based normalized gain chronologically. The data is listed by: *laboratory/teaching style (N = number of students) percentage total grade allotted to laboratories plus recitation*. Lab styles are labeled by traditional (T), Real Time Physics (RT), combination RT and T (RT-T), developed at the University of Illinois (IL)[26], and locally written PER-informed (PER). The lecture teaching styles are labeled by traditional lecture (TL), PER-informed lecture (PERL) and Inquiry-based instruction (INQ). Honors sections are labeled with an *H*.

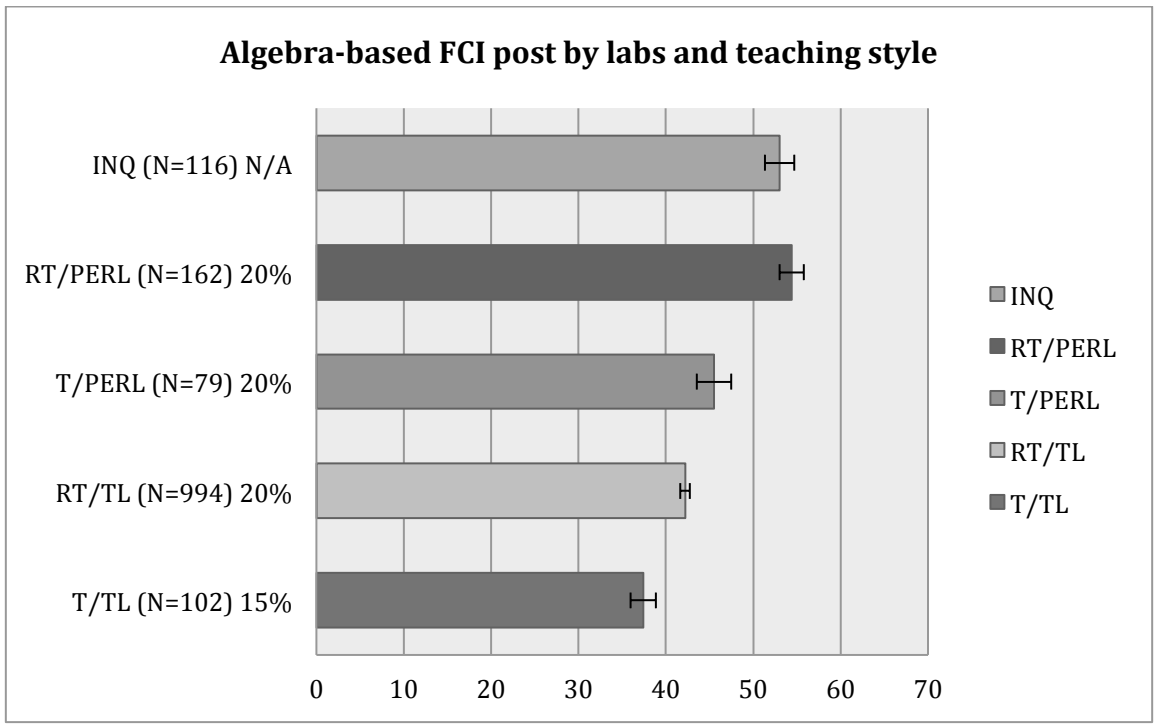

Figure 3. Algebra-based FCI post by laboratory and teaching style. The data is listed by: *laboratory/teaching style (N = number of students) percentage total grade allotted to laboratories plus recitation*. Lab styles are labeled by traditional (T), Real Time Physics (RT), combination RT and T (RT-T), developed at the University of Illinois (IL)[26], and locally written PER-informed (PER). The lecture teaching styles are labeled by traditional lecture (TL), PER-informed lecture (PERL) and Inquiry-based instruction (INQ). Honors sections are labeled with an *H*.

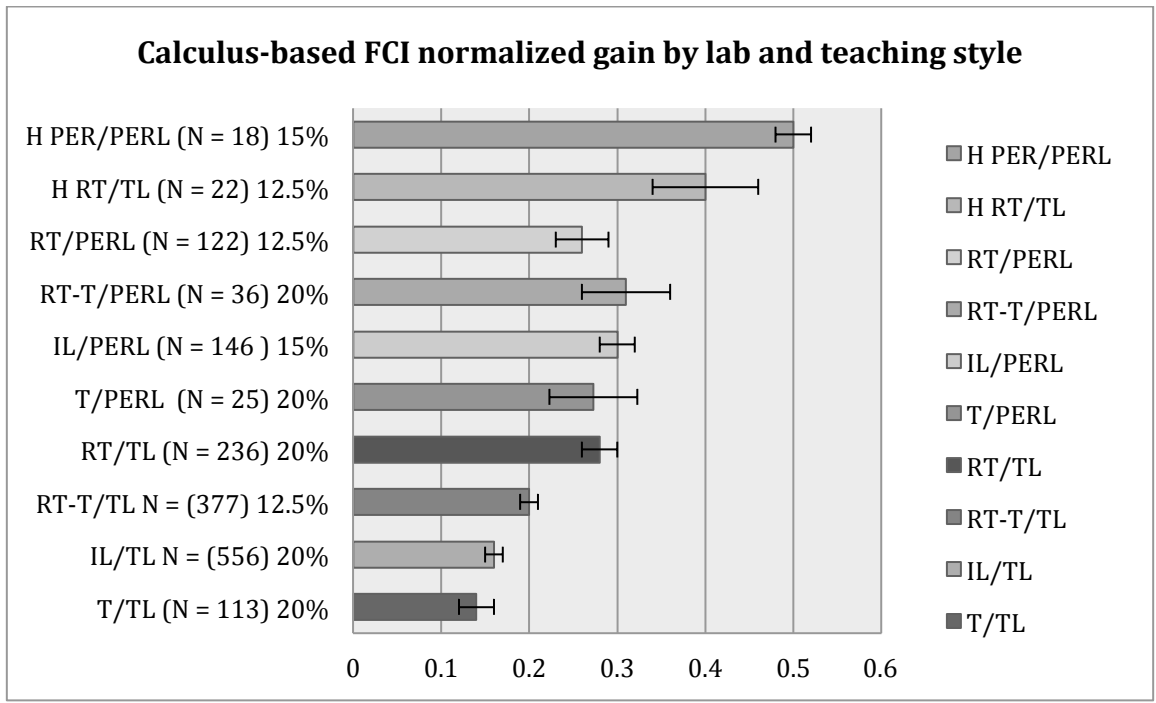

Figure 4. Calculus-based FCI normalized gain by laboratory and teaching style. The data is listed by: *laboratory/teaching style (N = number of students) percentage total grade allotted to laboratories plus recitation.* Lab styles are labeled by traditional (T), Real Time Physics (RT), combination RT and T (RT-T), developed at the University of Illinois (IL)[26], and locally written PER-informed (PER). The lecture teaching styles are labeled by traditional lecture (TL), PER-informed lecture (PERL) and Inquiry-based instruction (INQ). Honors sections are labeled with an *H*.

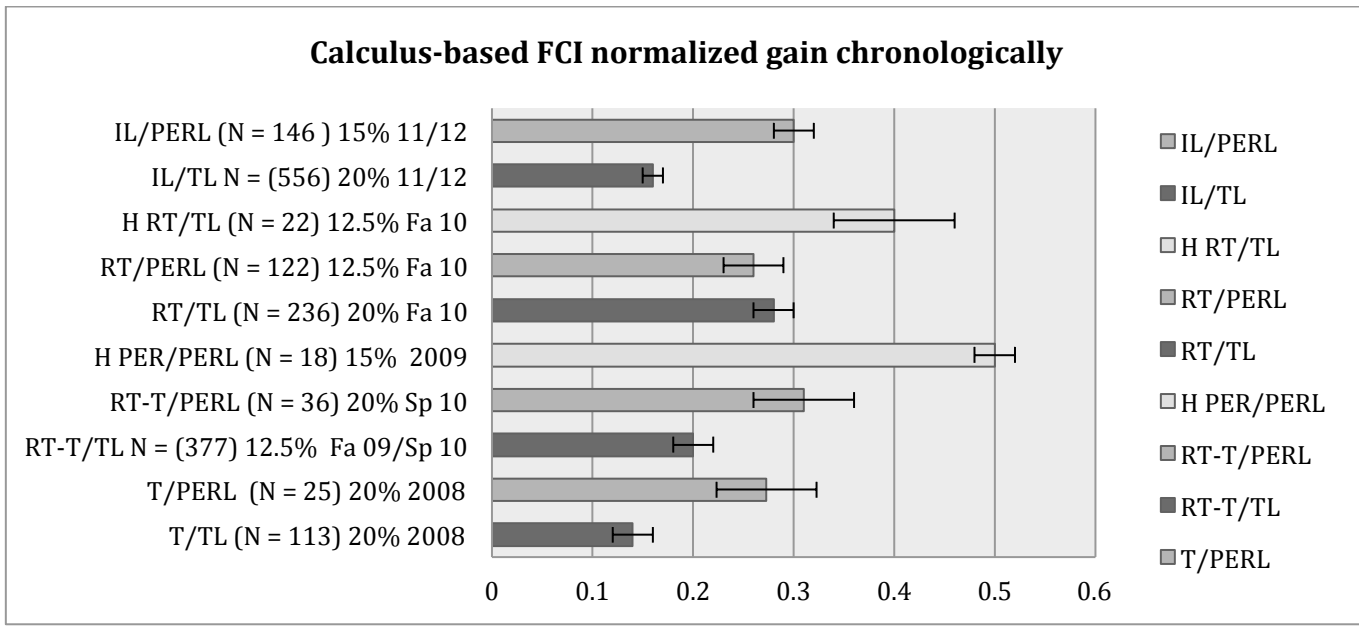

Figure 5. Calculus-based normalized gain chronologically. The data is listed by: *laboratory/teaching style (N = number of students) percentage total grade allotted to laboratories plus recitation.* Lab styles are labeled by traditional (T), Real Time Physics (RT), combination RT and T (RT-T), developed at the University of Illinois (IL)[26], and locally written PER-informed (PER). The lecture teaching styles are labeled by traditional lecture (TL), PER-informed lecture (PERL) and Inquiry-based instruction (INQ). Honors sections are labeled with an *H*.

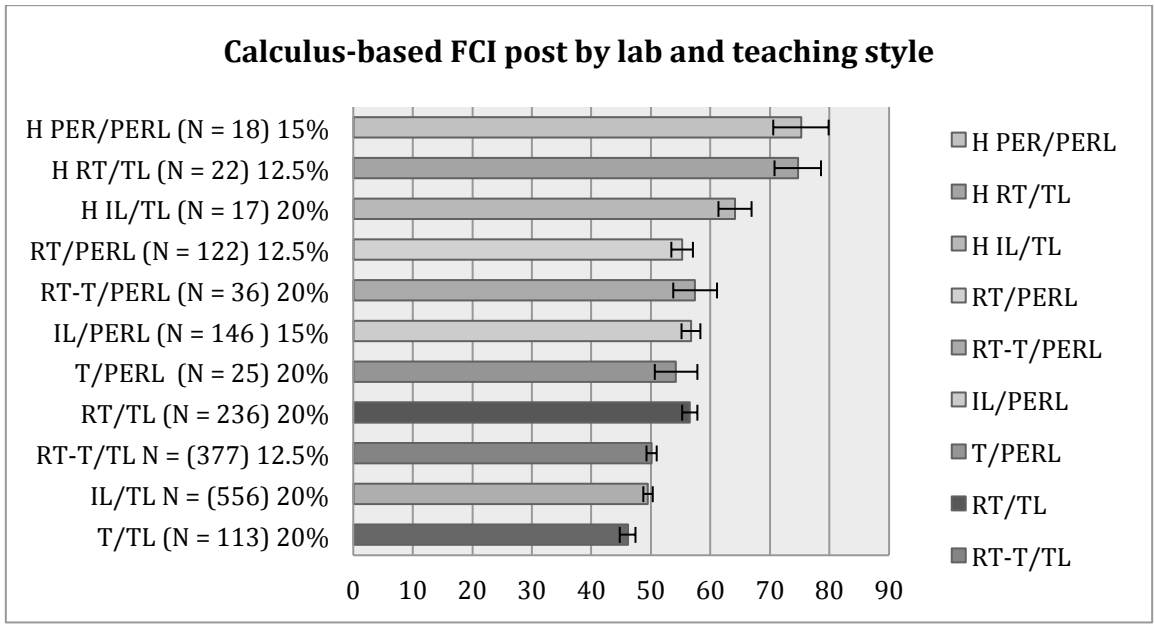

Figure 6. Calculus-based FCI post by laboratory and teaching style. The data is listed by: *laboratory/teaching style (N = number of students) percentage total grade allotted to laboratories plus recitation*. Lab styles are labeled by traditional (T), Real Time Physics (RT), combination RT and T (RT-T), developed at the University of Illinois (IL)[26], and locally written PER-informed (PER). The lecture teaching styles are labeled by traditional lecture (TL), PER-informed lecture (PERL) and Inquiry-based instruction (INQ). Honors sections are labeled with an *H*.

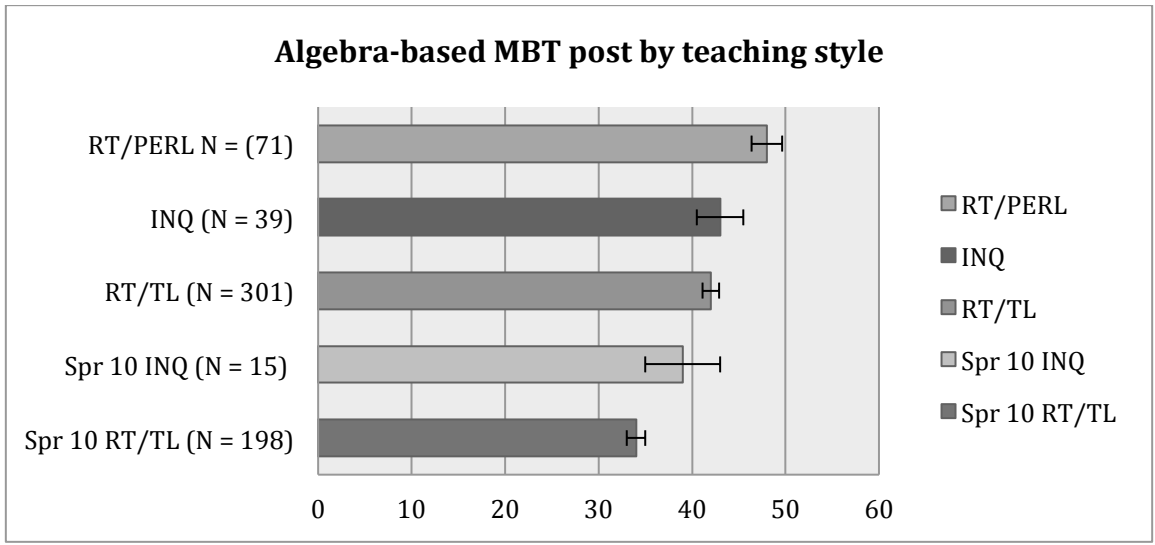

Figure 7. Algebra-based MBT post by lab and teaching style. The data is listed by: *laboratory/teaching style (N = number of students) percentage total grade allotted to laboratories plus recitation.* Lab styles are labeled by traditional (T), Real Time Physics (RT), combination RT and T (RT-T), developed at the University of Illinois (IL)[26], and locally written PER-informed (PER). The lecture teaching styles are labeled by traditional lecture (TL), PER-informed lecture (PERL) and Inquiry-based instruction (INQ). Honors sections are labeled with an *H*.

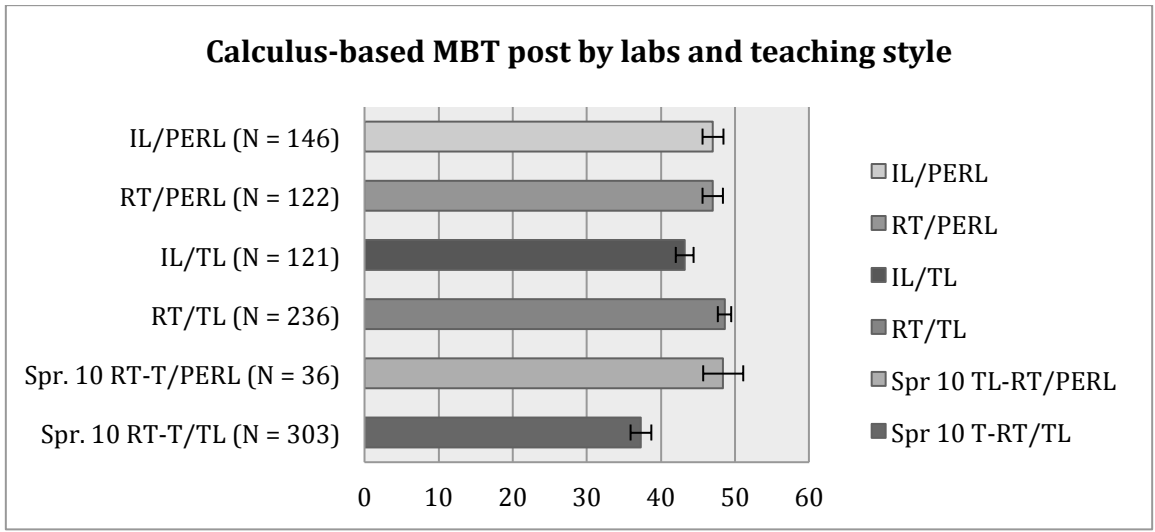

Figure 8. Calculus-based MBT post by lab and teaching style. The data is listed by: *laboratory/teaching style (N = number of students) percentage total grade allotted to laboratories plus recitation*. Lab styles are labeled by traditional (T), Real Time Physics (RT), combination RT and T (RT-T), developed at the University of Illinois (IL)[26], and locally written PER-informed (PER). The lecture teaching styles are labeled by traditional lecture (TL), PER-informed lecture (PERL) and Inquiry-based instruction (INQ). Honors sections are labeled with an *H*.

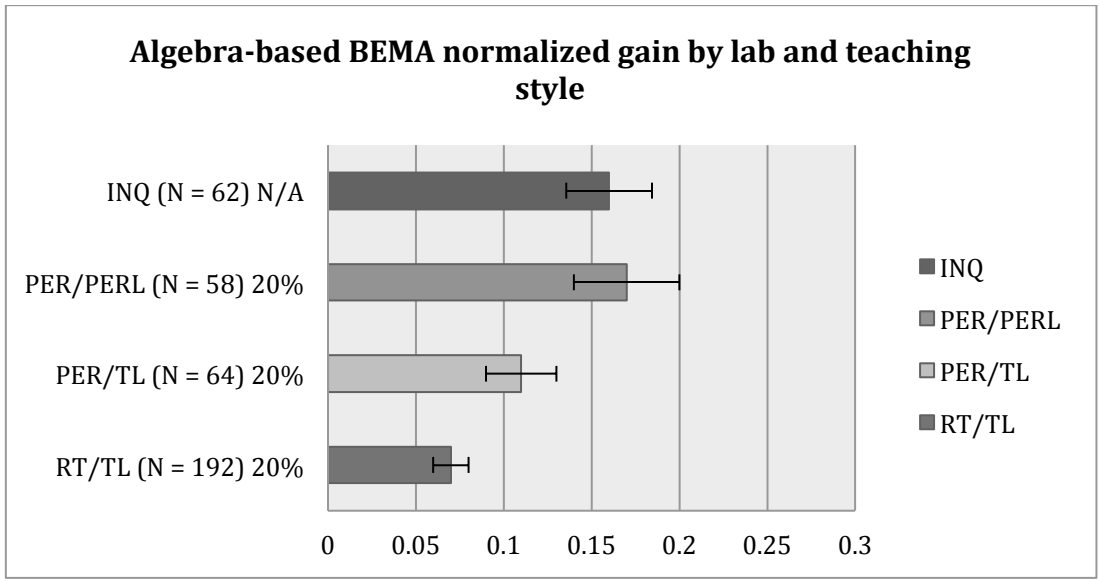

Figure 9. Algebra-based BEMA normalized gain by lab and teaching style. The data is listed by: *laboratory/teaching style (N = number of students) percentage total grade allotted to laboratories plus recitation.* Lab styles are labeled by traditional (T), Real Time Physics (RT), combination RT and T (RT-T), developed at the University of Illinois (IL)[26], and locally written PER-informed (PER). The lecture teaching styles are labeled by traditional lecture (TL), PER-informed lecture (PERL) and Inquiry-based instruction (INQ). Honors sections are labeled with an *H*.

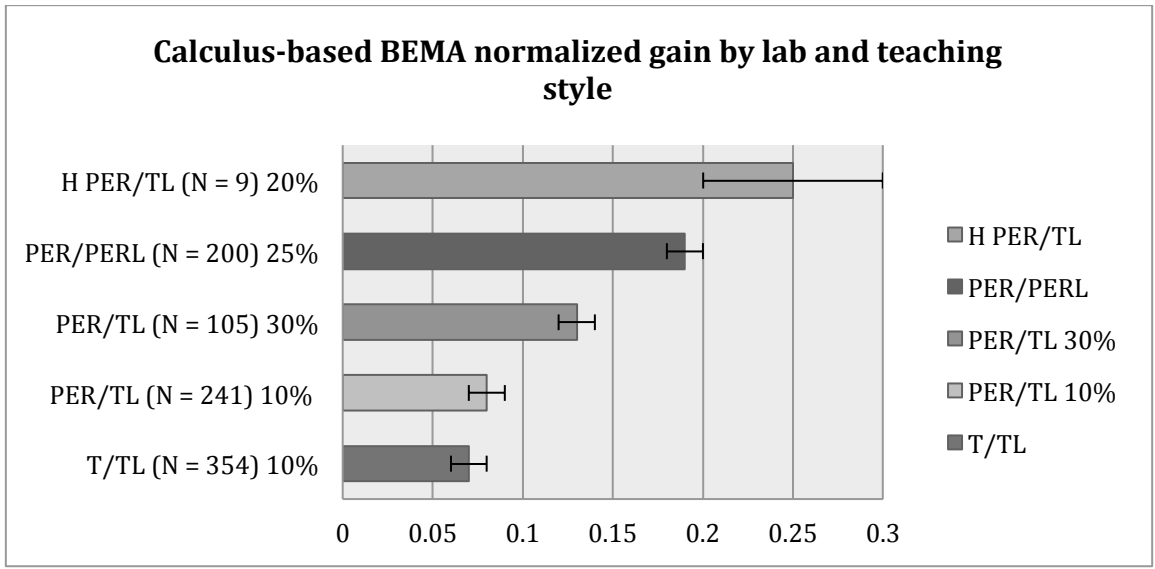

Figure 10. Calculus-based BEMA normalized gain by lab and teaching style. The data is listed by: *laboratory/teaching style (N = number of students) percentage total grade allotted to laboratories plus recitation.* Lab styles are labeled by traditional (T), Real Time Physics (RT), combination RT and T (RT-T), developed at the University of Illinois (IL)[26], and locally written PER-informed (PER). The lecture teaching styles are labeled by traditional lecture (TL), PER-informed lecture (PERL) and Inquiry-based instruction (INQ). Honors sections are labeled with an *H*.

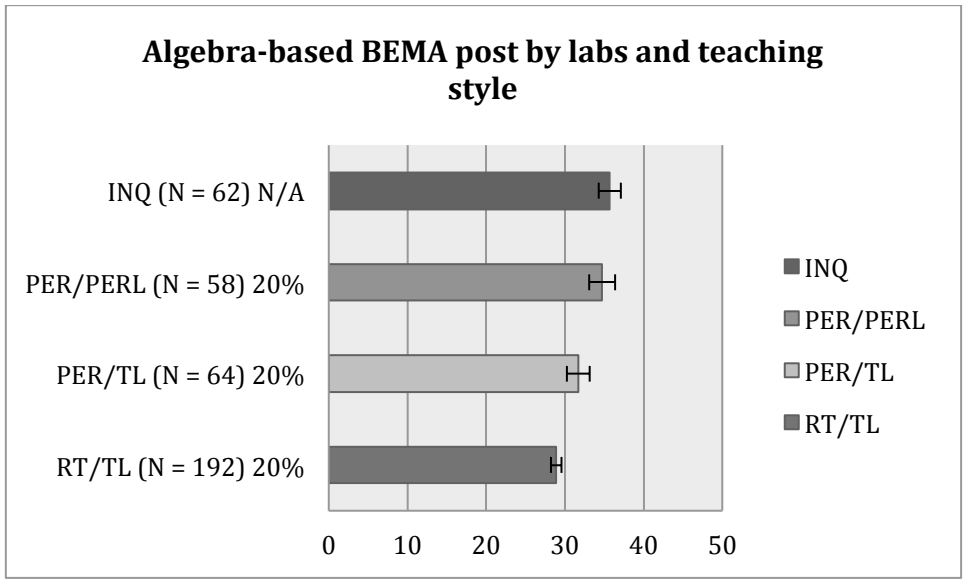

Figure 11. Algebra-based BEMA post-test by lab and teaching style. Data with an asterisk includes Spring 2011 data. The data is listed by: *laboratory/teaching style (N = number of students) percentage total grade allotted to laboratories plus recitation.* Lab styles are labeled by traditional (T), Real Time Physics (RT), combination RT and T (RT-T), developed at the University of Illinois (IL)[26], and locally written PER-informed (PER). The lecture teaching styles are labeled by traditional lecture (TL), PER-informed lecture (PERL) and Inquiry-based instruction (INQ). Honors sections are labeled with an *H*.

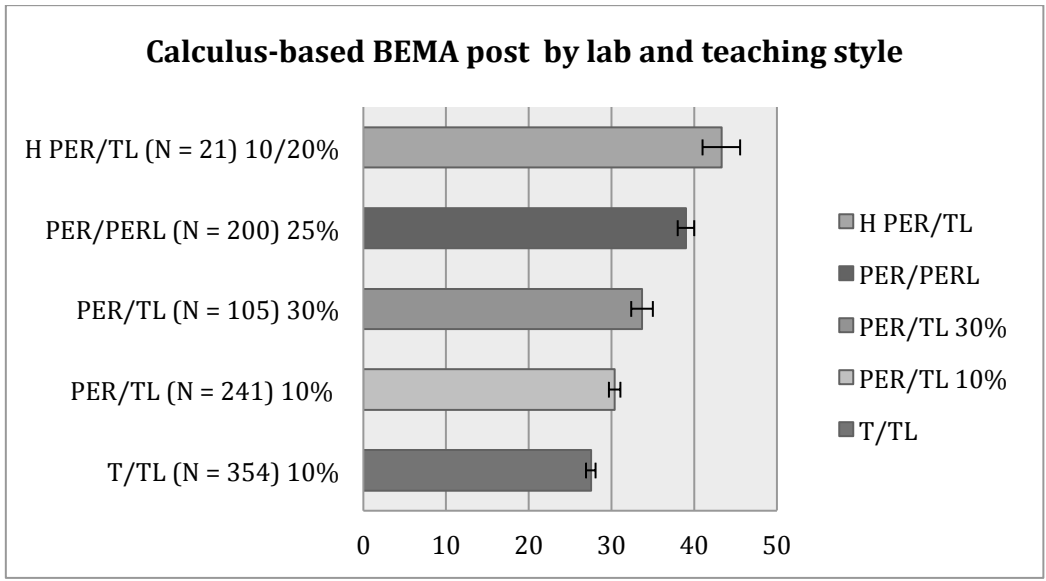

Figure 12. Calculus-based BEMA normalized gain by lab and teaching style. The data is listed by: *laboratory/teaching style (N = number of students) percentage total grade allotted to laboratories plus recitation*. Lab styles are labeled by traditional (T), Real Time Physics (RT), combination RT and T (RT-T), developed at the University of Illinois (IL)[26], and locally written PER-informed (PER). The lecture teaching styles are labeled by traditional lecture (TL), PER-informed lecture (PERL) and Inquiry-based instruction (INQ). Honors sections are labeled with an *H*.

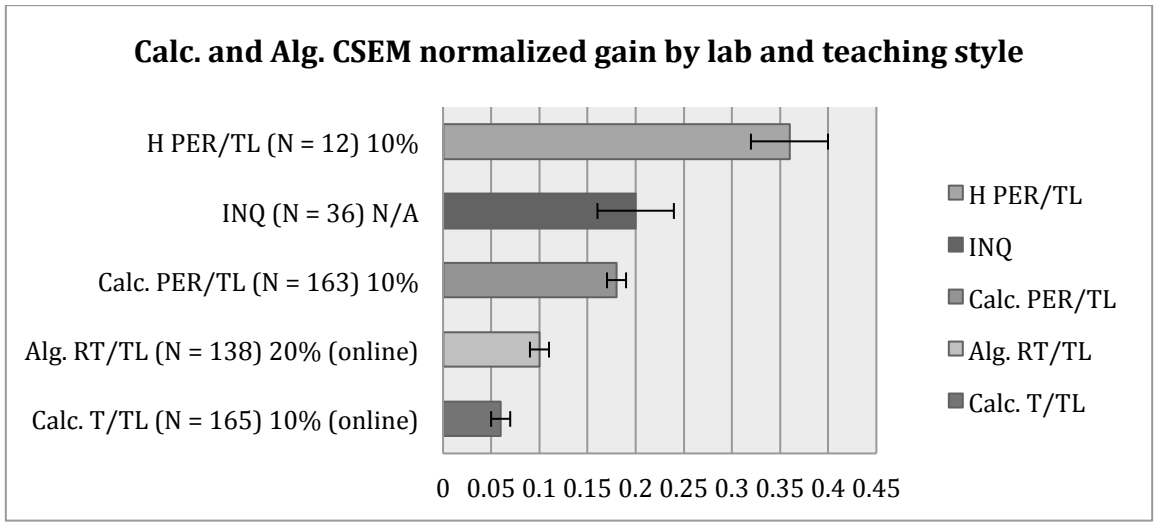

Figure 13. Calculus and algebra-based CSEM normalized gain by lab and teaching style. The data is listed by: *laboratory/teaching style (N = number of students) percentage total grade allotted to laboratories plus recitation*. Lab styles are labeled by traditional (T), Real Time Physics (RT), combination RT and T (RT-T), developed at the University of Illinois (IL)[26], and locally written PER-informed (PER). The lecture teaching styles are labeled by traditional lecture (TL), PER-informed lecture (PERL) and Inquiry-based instruction (INQ). Honors sections are labeled with an *H*.

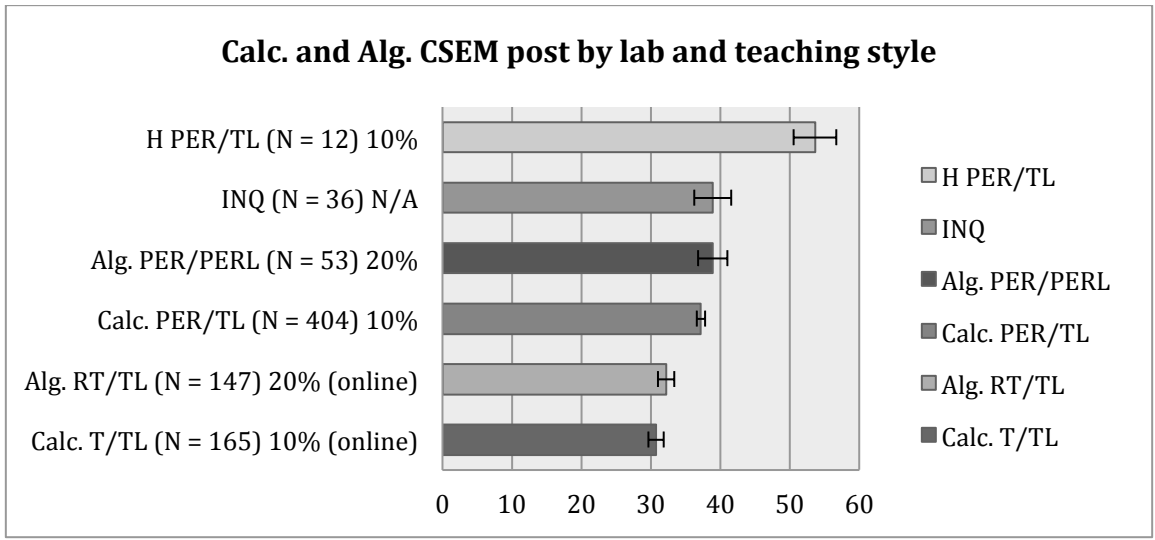

Figure 14. Calculus and algebra-based CSEM post by lab and teaching style. The data is listed by: *laboratory/teaching style (N = number of students) percentage total grade allotted to laboratories plus recitation.* Lab styles are labeled by traditional (T), Real Time Physics (RT), combination RT and T (RT-T), developed at the University of Illinois (IL)[26], and locally written PER-informed (PER). The lecture teaching styles are labeled by traditional lecture (TL), PER-informed lecture (PERL) and Inquiry-based instruction (INQ). Honors sections are labeled with an *H*.